\documentclass[aps,pra,preprint,superscriptaddress, amsmath,amssymp,showpacs]{revtex4}
\pdfoutput=1
\usepackage[T1]{fontenc}
\usepackage[latin1]{inputenc}
\usepackage[pdftex]{graphicx,color}
\usepackage{bm}        
\usepackage{amssymb}   
\usepackage{amsmath} 
\usepackage{color}
\def\jcp#1#2#3{J.~Chem.~Phys.~{\bf #1},\ #2\ (#3)}

\def\pra#1#2#3{Phys.~Rev.~A~{\bf #1},\ #2\ (#3)}
\def\prl#1#2#3{Phys.~Rev.~Lett.~{\bf #1},\ #2\ (#3)}

\begin{document}
\date{\today}
\flushbottom \draft
\title{Adiabatic Channel Capture Theory Applied to Cold Atom-Molecule Reactions: Li + CaH $\to$ LiH + Ca at 1 K} 

\author{Timur V. Tscherbul}
\affiliation{Chemical Physics Theory Group, Department of Chemistry, and Center for Quantum Information and Quantum Control, University of Toronto, Toronto, Ontario, M5S 3H6, Canada}\email[]{ttscherb@chem.utoronto.ca}
\author{Alexei A. Buchachenko}
\affiliation{Department of Chemistry, M.~V.~Lomonosov Moscow State University, Moscow 119991, Russia}\email[]{alexei@classic.chem.msu.su}
\affiliation{Institute of Problems of Chemical Physics RAS, Chernogolovka, Moscow District 142432, Russia}

\begin{abstract}
We use quantum and classical adiabatic capture theories to study the chemical reaction Li~+~CaH~$\to$~LiH~+~Ca. Using a recently developed {\it ab initio} potential energy surface, which provides an accurate representation of long-range interactions in the entrance reaction channel, we calculate the adiabatic channel potentials by diagonalizing the atom-molecule Hamiltonian as a function of the atom-molecule separation. The resulting adiabatic channel potentials are used  to calculate both the classical and quantum  capture probabilities as a function of collision energy, as well as the temperature dependencies of the partial and total reaction rates. The calculated reaction rate agrees well with the measured value at 1 K [V. Singh {\it et al.}, \prl {108}{203201}{2012}],  suggesting that the title reaction proceeds without an activation barrier. The calculated classical adiabatic capture rate agrees well with  the quantum result in the multiple partial wave regime of relevance to the experiment. Significant differences are  found only in the ultracold limit ($T<1$~mK), demonstrating that adiabatic capture theories can  predict the reaction rates with nearly quantitative  accuracy  in the multiple partial wave regime.
\end{abstract}

\maketitle
\clearpage
\newpage

\section{Introduction}


The high level of control over molecular degrees of freedom achieved in experiments with cold molecular ensembles is projected to make a profound impact on chemical physics \cite{ColdMoleculeBook,JunARPC, spectroscopy, Roman10,Roman08,njp09,molphys13,Roman13,John14,ChemRevBas,ChemRevJohn,ChemRevGoulven,KRb08,KRb10a,KRb10b,prl06,jcp07,Roman08}, particularly in the areas of chemical reaction dynamics \cite{ColdMoleculeBook,JunARPC, Roman08,njp09,molphys13}, molecular spectroscopy \cite{spectroscopy}, and energy transfer in molecular aggregates  \cite{Roman10,Roman13}. In particular, recent advances in magnetic trapping, buffer-gas cooling, and Stark deceleration of molecular radicals \cite{John14,JunARPC, ChemRevBas,ChemRevJohn} along with the realization of a dipolar quantum gas of KRb molecules in the ground rovibrational state \cite{KRb08,KRb10a,KRb10b} have opened the door to novel, previously unaccessible regimes of chemical reaction dynamics  characterized by quantum mechanical tunnelling, single partial-wave scattering, and quantum statistics. In these regimes, the effects of external  electromagnetic fields on the reactants' energy levels can greatly exceed the kinetic energy of their relative motion, which makes it possible to use the fields to control the  outcome of inelastic and reactive collisions \cite{ColdMoleculeBook,Roman08,prl06, JunARPC} (See Ref. \citenum{molphys13}  for a recent overview of  the field containing 853 references).


Measurements of ultracold chemical reaction rates were carried out in a dipolar gas of KRb molecules and K + KRb mixtures \cite{KRb10a,KRb10b} prepared in their absolute ground states by photoassociation of ultracold atoms \cite{KRb08}. The chemical reactions KRb + KRb and KRb~+~K were observed by measuring the loss of the reactants from an optical dipole trap at temperatures as low as 50 nK \cite{KRb10b}. Accurate quantum reactive scattering calculations based on {\it ab initio} potential energy surfaces were  carried out on the chemical reactions  Li~+~Li$_2$, Na~+~Na$_2$, and K + K$_2$  in the absence of external fields \cite{Pavel1,GoulvenK3} and on the reaction LiF~+~H~$\to$ Li~+~HF  in an electric field \cite{ourLiHF}. However, the computational cost of these calculations grows rapidly with increasing density of rovibrational states of the reaction complex, making it next to impossible to carry out converged calculations on  asymmetric reactions involving heavy atoms (such as Rb + K$_2$ $\to$ KRb + K), especially if the goal is to study the effects of  hyperfine interactions and external electromagnetic fields on low-temperature  reaction rates~\cite{Bohn10}.

To describe the experimental observations, Quem{\'e}n{\'e}r and Bohn combined the classical Langevin capture theory with quantum Wigner threshold laws and used the resulting quantum threshold (QT) model to calculate the dependence of KRb + KRb reaction rates  on the applied electric field \cite{Bohn10}, and the suppression effects due to external confinement  \cite{Bohn12} obtaining good qualitative agreement with experiments \cite{KRb10a,KRb10b}.
 Idziaszek and Julienne developed a more sophisticated model based on multichannel quantum defect theory (MQDT) with a complex potential to describe the loss dynamics at short range \cite{Julienne10}. By introducing a dimensionless quantum defect parameter $y$, they were able to describe a continuum of  physical scenarios ranging from complete ($y=1$) to non-existent ($y=0$) absorption of reactive flux at short range. The MQDT model reduces to the QT model in the limit  $y=1$ (also known as the capture approximation \cite{Langevin} and the ``universal'' limit), and has been applied to describe the KRb experiments in the single partial-wave regime  ($L=0$ or 1) \cite{Julienne10}. More recently, Jachymski {\it et al.} extended the model to long-range polarization potentials and higher partial-waves and obtained good agreement with recent experimental measurements of Penning ionization rates in cold Ar + He$^\star$ and H$_2$ +  He$^\star$ collisions over a wide range of temperatures (10 mK - 30 K) using a single adjustable parameter $y=0.007$ \cite{Julienne13,Narevicius}.
Bo Gao derived analytical expressions for MQDT capture probabilities and reaction rates for arbitrary $L$ and demonstrated that in the high-temperature limit, the rate coefficients follow a universal behaviour $K \propto T^{1/6}$ predicted by the classical Gorin model  \cite{BoGao}.

The QT and MQDT models \cite{Bohn10,Julienne10,Julienne13,BoGao} provide simple analytical expressions for capture probabilities and reaction rates for the interaction potentials displaying the generic asymptotic behavior $\propto R^{-n}$, where $n=4$ or 6 are appropriate for the field-free ion-neutral or neutral-neutral reactions in the limit  $T \to 0$ (here, $R$ is the distance between the reactants' centers of mass). In the last 40 years, more sophisticated classical   \cite{ClaryARPC90,TroeACP97} and quantum \cite{ClaryMP82,RackhamJCP03,RackhamJCP04} adiabatic channel (AC) capture theories were developed and applied to calculate the reaction rates.  The AC capture theories (which we will also refer to as  ``adiabatic capture theories'') use $R$-dependent eigenvalues of the Hamiltonian of the reaction complex in the entrance reaction channel (the adiabatic channels) as a starting point to compute the capture probabilities. As a result, these theories can describe  capture by anisotropic potential energy surfaces (PESs), which is crucial for the proper description of reaction dynamics at elevated temperatures. At ultralow temperatures,  the predictions of the quantum AC capture theory agree well with those of the universal QT and MQDT models \cite{Julienne10}, as demonstrated by recent calculations of KRb~+~KRb reaction rates based on a highly accurate KRb-KRb long-range potential obtained from correlated {\it ab initio} calculations \cite{VMU5,ourKRb}.

A wide variety of molecules of chemical interest (such as OH, NH, CaH,  ND$_3$, C$_8$H$_{10}$, and C$_6$H$_5$CN) have been produced at low temperatures (1 mK - 1 K) via buffer-gas cooling and Stark deceleration techniques \cite{Roman08,molphys13}. Collisions and reactions in such cold (as opposed to ultracold) molecular gases are determined by multiple partial wave scattering. Of particular interest is the crossover regime between the purely quantum ($L=0$) and  semiclassical (many-$L$) atom-molecule reaction dynamics \cite{Bohn10,Julienne10,BoGao} in which theoretical predictions  have not  been tested due to the lack of experimental data. In a recent experiment, the group of Jonathan Weinstein at the University of Nevada, Reno has measured the rate of the Li~+~CaH $\to$ LiH + Ca reaction by detecting the loss of buffer-gas cooled CaH reactants and the appearance of LiH products \cite{CaHLiJonathan2012}. They found that the reaction occurs rapidly, with a rate coefficient of $3.6\times 10^{-10}$ cm$^3$/s at 1 K.  At this temperature, up to a dozen partial waves contribute to the reaction probability; thereby providing a unique opportunity to test the applicability of the capture models in a previously unexplored temperature regime. 
 
 Here, we apply the adiabatic capture theory to study the dynamics of the Li + CaH chemical reaction in the temperature range from 10 nK to 100 K. To calculate the ACs, we use an accurate {\it ab initio} potential energy surface for the triplet $^3A'$ state of the Li-CaH reaction complex  \cite{ourCaHLi}, which has the same asymptotic properties as the ground electronic state of  $^1A'$ symmetry and provides an accurate description of long-range interactions in the entrance reaction channel. We study the crossover between the quantum and classical Langevin regimes, and compare the results at 1 K with the experimental observations. 
Our calculated quantum and classical adiabatic capture rates are  $K = 7.1$ and $7.2 \times 10^{-10}$ cm$^3$/s, respectively, in good agreement with the experimental value ($3.6 \times 10^{-10}$ cm$^3$/s, within a factor of 2 uncertainty \cite{CaHLiJonathan2012}). These results 
suggest that the Li~+~CaH chemical reaction proceeds without an activation barrier and its low-temperature dynamics is accurately described by a classical AC capture theory, providing the first validation of the theory for neutral atom-molecule chemical reactions at low temperatures beyond the single-partial-wave regime.


The rest of this article is organized as follows. In Sec. II, we outline the basics of the  AC capture theory and the procedures for calculating the ACs and the corresponding capture probabilities. Sec. III presents the results of our calculations of quantum and classical capture probabilities, cross sections, and rate coefficients, and compares the results with experiment \cite{CaHLiJonathan2012}. Sec. IV presents our conclusions and outlines possible directions in which this work can be extended. Atomic units are used throughout unless otherwise noted.

\section{Theory}


\subsection{Potential energy surface and adiabatic channels}

The interaction between ground-state Li($^2$S) atoms and CaH($X^2\Sigma^+$) molecules in the entrance reaction channel gives rise to the singlet and triplet electronic states of $^1A'$ and $^3A'$ symmetry within the $C_s$ point group. In the product channel, the  $^1A'$ state correlates to the Ca($^1$S) + LiH($X^1\Sigma^+$) asymptotic limit,  which lies $\Delta E\sim$0.7 eV below the reactants \cite{HuberHerzberg}, making the Li + CaH $\to$ LiH + Ca reaction strongly exothermic. The large amount of energy released in the reaction can lead to vibrationally hot LiH($v$) products with $v= 4-5$ \cite{CaHLiJonathan2012}.  The triplet state $^3A'$  correlates to the Ca($^1$S) + LiH($^3\Sigma^+$) product limit. Since the lowest $a^3\Sigma^+$ state of LiH is repulsive (see, {\it e.g.}, Ref.~\citenum{BotalibGadea}), this limit lies $\sim$1.7 eV above the Li + CaH$(v=0,j=0)$ asymptote and hence the reaction channel Li($^2$S) + CaH($X^2\Sigma^+$) $\to$ Ca($^1$S) + LiH($^3\Sigma^+$)  is closed at low collision energies considered in this work.

We have recently performed {\it ab initio} calculations of the $^3A'$ Li-CaH PES using the open-shell coupled cluster method with single and double excitations and non-iterative corrections to triple excitation [CCSD(T)] and a large quadruple-$\zeta$-quality correlation-consistent basis set augmented by diffuse and $R$-centered bond functions \cite{ourCaHLi}. We found that the Li-CaH collision complex has a binding energy of $\sim$0.88 eV, and that the chemical rearrangement Li + CaH $\to$ LiH + Ca on the triplet PES is forbidden by spin conservation, in agreement with the qualitative arguments given above. These arguments form the basis of a theoretical proposal for controlling  chemical reactions of open-shell reactants by inducing  spin-forbidden transitions with combined electric and magnetic fields  \cite{prl06,jcp07}. 


It is important to note that in the asymptotic reactant limit, the PESs of both the triplet $^3A'$ and the singlet $^1A'$ states are determined by the same (spin-independent) long-range induction and dispersion forces, whereas the spin-dependent exchange and chemical interactions come into play at much shorter distances ($R < 4-5$ {\AA} for alkali-metal atoms \cite{Smirnov,Stwalley}), well into the region where a collision complex is formed.  Because only the long-range part of the atom-molecule interaction is required to model the reaction within the framework of adiabatic capture theories  \cite{Julienne10,Bohn10}, we can model the Li + CaH chemical reaction using the highly accurate triplet PES calculated in Ref.~\cite{CaHLiJonathan2012}  instead of performing elaborate (and generally less accurate) multireference calculations of the singlet PES. 
Accordingly, we set the multiplicity factor, which accounts for the degeneracy of the triplet state to be equal to unity, the value appropriate for the $^1A'$ PES.


To describe the Li-CaH reaction complex, we  use the standard Jacobi coordinates $r$, $R$ and $\theta$, where $r$ is the Ca--H internuclear distance, $R$ is the distance between the center of mass of CaH and Li, and $\theta$ is the angle between the vectors {\bf r} and {\bf R}. In these coordinates,  the Hamiltonian of the collision complex can be written as
\begin{equation}
\label{H3D}
\hat{H} = -\frac{1}{2\mu R}\frac{\partial^2}{\partial R^2}R
          -\frac{1}{2mr}\frac{\partial^2}{\partial r^2}r
	  +\frac{{\bf L}^2}{2\mu R^2}
	  +\frac{{\bf j}^2}{2mr^2} +  V(r,R,\theta),
\end{equation}
where $m$ and $\mu$ stand for  the reduced masses of CaH and Li-CaH, respectively, {\bf L} is the orbital angular momentum of the colliding partners {\bf L} = {\bf J} - {\bf j} with {\bf J} being the total angular momentum and {\bf j} -- the rotational momentum of the CaH molecule. The spin-rotation coupling in the CaH molecule is small \cite{Shayesteh} and weakly affects collision dynamics (as a result, the spin depolarization in Li + CaH collisions is strongly suppressed \cite{ourCaHLi}). We also neglect vibrational motion and fix $r$ at the calculated equilibrium value for CaH, $r_e = 2.012$ \AA. The Hamiltonian (\ref{H3D}) is then reduced to 
\begin{equation}
\label{H2D}
\hat{H} = -\frac{1}{2\mu R}\frac{d^2}{dR^2}R
	  +\frac{{\bf L}^2}{2\mu R^2}
	  +B_e{\bf j}^2 +  V(r_e,R,\theta),
\end{equation}
with $B_e = 4.277$ cm$^{-1}$ ({\it cf.} the experimental value 4.228 cm$^{-1}$ \cite{Shayesteh}). 


The AC potentials are the key input to both the quantum and classical adiabatic capture models. To obtain the AC potentials, we  diagonalize, at fixed $R$ values, the matrix of the Hamiltonian (\ref{H2D}) in the symmetry-adapted rigid rotor function basis set \cite{GDB,Nadine}
\begin{eqnarray}
\Theta^{JMp}_{j\Omega}(\hat{{\bf r}} \cdot \hat{{\bf R}}) & = &
 \left[ \frac{2J+1}{8\pi(1+\delta_{\Omega0})} \right]^{1/2} \times \nonumber \\
 & &
\left[D^{\ast J}_{M\Omega}(\alpha,\beta,0)Y_{j\Omega}(\theta,\varphi)
 + pD^{\ast J}_{M\,-\Omega}(\alpha,\beta,0)
 Y_{j\,-\Omega}(\theta,\varphi) \right],  
\label{Theta}
\end{eqnarray}
where $D^{\ast J}_{M\Omega}(\alpha,\beta,0)$ is the Wigner $D$-function, which depends on the Euler angles connecting  the body-fixed frame related to {\bf R} to the space-fixed frame, in which the {\bf J} vector has the projections $M$ and $\Omega$, respectively, $Y_{j\Omega}(\theta,\varphi)$ is the spherical harmonic function describing the rotation of the CaH molecule in the body-fixed frame and $p = \pm 1$ is the inversion parity. We are interested in the lowest-energy channels correlating to the ground rovibrational  state of  CaH ($v=0,j=0$) in the limit $R\to \infty$, which can be labelled by the single quantum number $L$ [$L=J$, $\Omega = 0$, $p = +1$, $(2L+1)$-fold degeneracy in $M$].

For each AC potential $v_L(R)$, we calculate  the capture probability $P^L(E)$ as a function of collision energy $E$  using the different approaches as outlined in the following sections. 

\subsection{Classical capture theory} 

The Langevin capture probability  is given by \cite{Langevin}
\begin{equation}
P_\mathrm{C}^L(E) = \left\{\begin{array}{lcl} 0, & \mathrm{if} & E < v^*_L \\
                                                                 1, & \mathrm{if} & E \ge v^*_L. \end{array}\right.
\label{Pclass}
\end{equation}
where $v^{*}_L$ is the maximum of the AC potential $v_L$.

For $L=0$, $v^*_L = 0$ and the capture probability is equal to unity at any $E$. The capture cross section is related to the capture probability via 
\begin{equation}
\sigma(E) = \frac{\pi}{2\mu E}\sum_{L=0}^{\infty}(2L+1)P_\mathrm{C}^L(E),
\label{CSsigma}
\end{equation}
whereas the rate coefficient at temperature $T$ may be calculated from the capture cross section as
\begin{equation}
K(T) = \frac{1}{k_BT}\left(\frac{8}{\pi\mu k_BT}\right)^{1/2}\int_{0}^{\infty}\sigma(E)\exp(-E/k_BT)EdE,
\label{RCk}
\end{equation}
where $k_B$ is the Boltzmann constant. 

Combining the above three equations, we obtain the following expression for the classical capture rate coefficient \cite{RamillonMcCarrol,TroeCPL85} 
\begin{equation}
K_C(T) = \left(\frac{2\pi}{\mu^3k_BT}\right)^{1/2}\left[1 + \sum_{L=1}^{\infty}(2L+1)\exp(-v^*_L/k_BT)\right].
\label{RCclass}
\end{equation}
This expression establishes that the dominant contribution to the capture rate  at ultralow temperatures comes from the lowest partial wave, and diverges in the $T\to 0$ limit. To avoid this unphysical behavior, we use  the quantum capture theory described below.

\subsection{Quantum capture theory}

 The basic ideas of the quantum capture theory were laid down by Vogt and Wannier \cite{Vogt-Wannier} and further developed in Refs.~\cite{ClaryMP82,RackhamJCP03,Light}, and in subsequent theoretical work on the capture problem for generic $R^{-n}$ long-range potentials, which reformulated the classical concept of the ``dividing surface'' (that separates the configuration space of the system into the reactants and reaction complex regions)    in terms of the boundary conditions on the complex's wavefunction. This reformulation allows one to consider capture theory on the same grounds as the conventional transition state theory \cite{Fernandez-Ramos}. In the reactant limit ($R \to \infty$) the wavefunction of the reaction complex is represented as a combination of incident and (elastically) scattered waves
\begin{equation}
\psi_E(R) \to \exp{(-\imath kR)} + A(E)\exp{(\imath kR)},
\label{WFtoinfty}
\end{equation}
whereas in the collision complex limit ($R \to 0$), the wavefunction is given by the transmitted wave:
\begin{equation}
\psi_E(R) \to \alpha(E)\exp{(-\imath kR)},
\label{WFtozero}
\end{equation}
with $k = (2\mu E)^{1/2}$ being the wavevector for the collision. Equation (\ref{WFtozero}) neglects the probability flux from the reaction complex region back to the reactants, thereby stating that the probability of finding the system in the collision complex region is equal to the reaction probability. For an AC characterized by the quantum number $L$, the quantum capture probability is equal to the transmission coefficient
\begin{equation}
P_\mathrm{Q}^L(E) = T_L(E) = |\alpha(E)|^2.
\label{Pquant}
\end{equation}

This standard and transparent formulation meets, however, with a singularity problem. For the $R^{-n}$ long-range case, the problem reflects the singularity of the potential in the limit $R \to 0$. Thorough analytical treatments of the singular potential scattering problem  led to the effective range and quantum defect theories for capture by inverse-power potentials in the limit $E \to 0$ (see, {\it e.g.} Refs.~\cite{Julienne10,Julienne13,BoGao,OMalley,Watanabe,Cavangero,Gao1998,Gao2009,Michelli}.)
While the adiabatic capture models of chemical reactions are based on  non-singular PESs,  they can exhibit the singularity problem if expressed in Jacobi coordinates, since the reactant's Jacobi coordinates are not suitable for the description of the reaction complex and of the product channel. As a result of this coordinate problem \cite{PP87}, as $R$ decreases, the system encounters a repulsive potential wall rather than turns into a  product valley. Because full reflection from the repulsive wall is incompatible with the capture approximation, Eq. (\ref{WFtozero}) should be applied at a certain point $R_0$, which lies far enough into the collision complex region, yet not too close to the origin
\begin{equation}
\psi_E(R_0) = \alpha(R_0;E)\exp{(-\imath k_0R_0)},
\label{WFtoR0}
\end{equation}
with $k_0=\{2\mu[E-v_L(R_0)]\}^{1/2}$. In the absence of any information about the reaction complex, the choice of $R_0$, while   subject to the above constraints, is otherwise arbitrary. However, there exists a range of $R_0$ where the transmisson coefficient $|\alpha(R_0; E)|^2$ oscillates weakly \cite{VMU5,ourKRb}. Averaging over this range increases the accuracy of the numerical capture probabilities \cite{ourKRb}. Once $P_Q^L$ are known, the capture cross sections and rate coefficient can be obtained using Eqs.(\ref{CSsigma}) and (\ref{RCk}). 

Equations (\ref{WFtoinfty}) and (\ref{WFtozero}) picture the capture process as a combination of barrier transmission and reflection effects. To account for the former, it is easy to implement the standard Wentzel-Kramers-Brillouin (WKB) correction, which replaces the classical capture probability at  $E < v^*_L$ by \cite{LL,Froman} 
\begin{equation}
P_\mathrm{WKB}^L(E) = \exp{\left[ -2\int_{R^L_-}^{R^L_+}k^L(R)dR \right] },
\label{PWKB}
\end{equation}
where $k^L(R)=\{2\mu[E-v_L(R)]\}^{1/2}$ and $R^L_-$, $R^L_+$ are the locations of its zeroes. The effect of overbarrier reflection is neglected leaving classical probability definition (\ref{Pclass}) for $E \ge v^*_L$ unaltered.

\section{RESULTS}

The AC potentials were calculated numerically on a dense radial grid from $R=4$ {\AA}  to $R=5000$~{\AA}. The {\it ab initio} points were  smoothly extrapolated to an analytical  long-range expansion in inverse powers of $R$ beyond 18 {\AA}. The basis functions ({\ref{Theta}) with $j\le 20$ were used in the variational calculations to ensure the convergence of the ACs within 0.01 \%. Test calculations indicate that the partial wave summation in Eq.(\ref{CSsigma}) with  $L \le 60$ is sufficient to converge the rate coefficients at $T \leq 100$ K. 

Figure~\ref{ACpots} shows the AC potentials for $L = 0, 20, 40$ and 60. The long-range fit to the $v_0(R)$ potential gives the dispersion coefficient at the leading $R^{-6}$ term $C_6 = 1480 \pm 10$ a.u. The corresponding effective potentials (EPs) 
\begin{equation}
v_L(R) = \frac{L(L+1)}{2\mu R^2} - \frac{C_6}{R^6}
\label{DispPot}
\end{equation}
are also presented in Fig.~\ref{ACpots} by dashed lines. The true adiabatic channel potentials are remarkably more attractive due to the higher-order long-range interaction components. As a result, they have lower barriers shifted towards longer distances with respect to purely dispersion EPs. The difference magnifies with $L$, as illustrated in the insets in Fig.~\ref{ACpots}. 

\begin{figure}[t]
\includegraphics[width=0.8\textwidth]{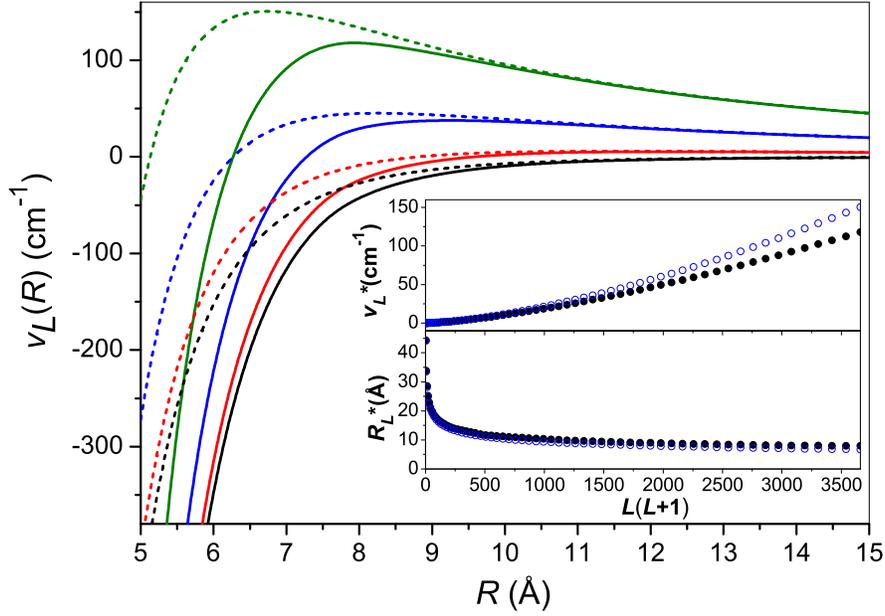}
\caption{Adiabatic channels (solid lines) and effective potentials (dashed lines) for $L = 0, 20, 40$ and 60. Insets show the variations of barrier position $R^*_L$ and height $v^*_L$ with $L(L+1)$ for AC (filled circles) and EP (open circles).} 
\label{ACpots}
\end{figure}

To calculate the quantum capture probabilities, we used the finite-difference Truhlar-Kuppermann method \cite{TK}, adapted as described in Ref.~\cite{VMU5}. The main advantage of this method over the propagation-based techniques is the possibility of flux conservation control. In all calculations the radial grids were used in such a way that the sum of reflection and transmission coefficients, $|A|^2 + |\alpha|^2$ in Eqs~(\ref{WFtoinfty}) and (\ref{WFtoR0}) differs from unity by less than 10$^{-4}$. At the lowest energies, the grids consisted up to 80,000 points and extended out to 5000 {\AA}. The $R_0$ dependence of the capture probability  was investigated carefully for different $L$ and $E$ and two ranges of $R_0$ were identified where the deviations of the transmission coefficients from their average values do not exceed 30~\% ($R_0\in [8-12]$ and [19-26] \AA). The final result was taken as an average over 8 $R_0$ points within these ranges~\cite{VMU5}. The capture probabilities were computed on a dense grid of collision energies $E$ from 10$^{-16}$ to 1000 a.u. (no calculations were performed at low- or high-energy limits if the deviations from  0 or 1 are found to be negligible). The capture cross sections were evaluated by summing up the partial contributions with $L \le 20$. 

\begin{figure}[t]
\centering
\includegraphics[width=0.5\textwidth]{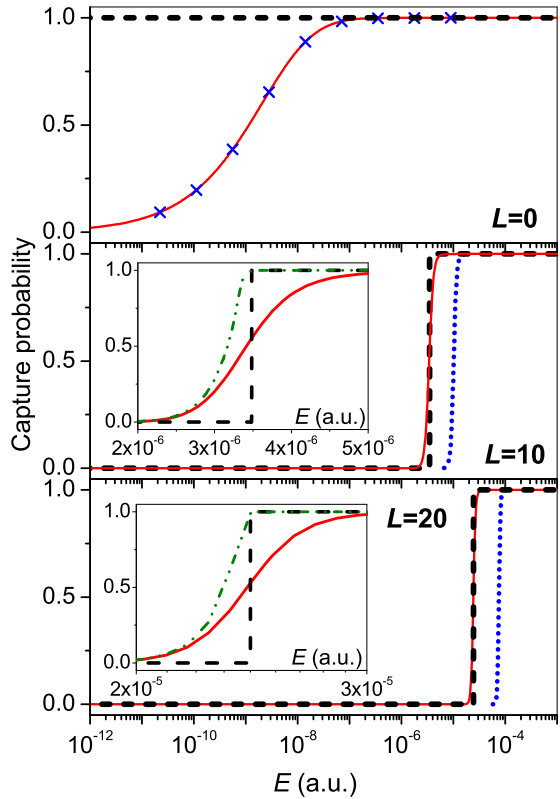}
\caption{Quantum (solid lines) and classical (bold dashed lies) capture probabilities calculated for the AC potentials and $L=0$, 10 and 20 (top, middle and bottom panels, respectively).  Crosses or dotted lines correspond to quantum calculations within the EP model. The insets expand the probability fall-off regions and show WKB-corrected classical probabilities by dash-dotted lines.} 
\centering
\label{QProbs}
\end{figure}

Figure~\ref{QProbs} shows the results obtained within the AC and EP models at selected $L$ values. In the purely attractive $L=0$ case (top panel) the capture probability slowly declines from one to zero as the collision energy decreases by $\sim$7 orders of magnitude. No difference is observed between the calculations with AC and EP potentials. As the centrifugal barrier emerges and grows up with $L$, the probability falls off much faster, more and more resembling the classical step function. The difference between the AC and EP results, which becomes evident in this regime, reflects the difference in position and shape of the centrifugal barrier ({\it cf.} Fig.~\ref{ACpots}). The insets in the middle and bottom panels provide an enlarged view of the capture probability fall-off for the AC potentials. The close similarity in the behavior of the probabilities for $L=10$ and 20 is consistent with previous theoretical work, which recommended using general analytical fitting functions for capture probabilities for the polarization ($R^{-4}$) and dispersion ($R^{-6}$) potentials (see, {\it e.g.}, Refs.~\cite{Klots1976,Dashevskaya2003,Nikitin2010}). The insets also show the WKB-corrected classical probabilities, which provide an upper bound for both the classical and quantum results. 

\begin{figure}[t]
\centering
\includegraphics[width=0.7\textwidth]{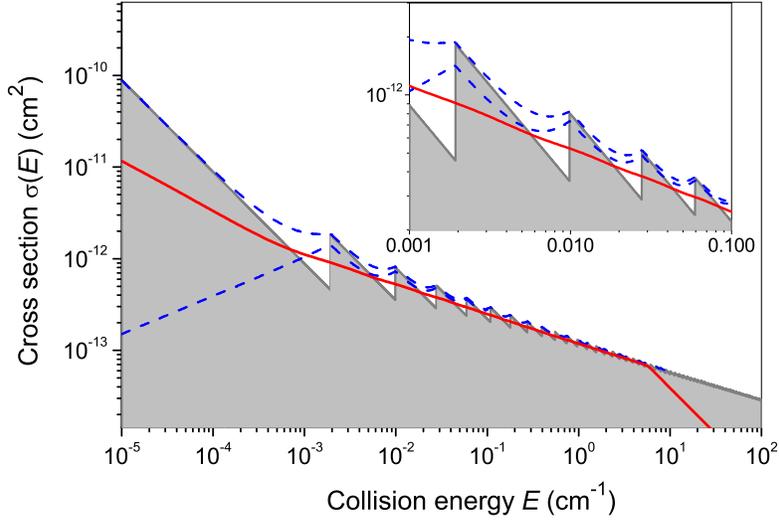}
\caption{Quantum (solid line), classical (shaded area) and WKB-corrected (dashed lines) capture cross sections. The WKB-corrected results are given with (upper line) and without (lower line) the $L=0$ contribution. The inset is an expanded view of the collision energy range corresponding to the onset of classical capture for $L = 1-4$. } 
\centering
\label{CS}
\end{figure}

 Figure~\ref{CS} shows the capture cross sections. The ``sawtooth-like'' classical dependence reflects the discrete probability definition (\ref{Pclass}), in part smoothed out by the WKB tunneling correction. As expected based on their definition (\ref{CSsigma}), both the classical and WKB-corrected cross sections increase  too fast  in the $E \to 0$ limit, $\sigma \propto E^{-1}$, whereas the quantum cross section exhibits a slower dependence $\propto E^{-1/2}$. This difference originates entirely from $s$-wave ($L=0$) scattering and features quantum reflection from a barrierless potential.  At $E> 10^{-3}$ cm$^{-1}$, close to the height of the $p$-wave centrifugal barrier, the quantum capture cross section approaches the classical result and the matching improves as the collision energy increases further as shown in the inset in Fig.~\ref{CS}. We note that the quantum cross section summed over the  lowest 21 partial waves is sufficient below $E \approx 5$ cm$^{-1}$. Above this energy, the classical capture cross section provides a good approximation to the quantum result, as shown in Fig.~\ref{CS}.

\begin{figure}[t]
\centering
\includegraphics[width=0.7\textwidth]{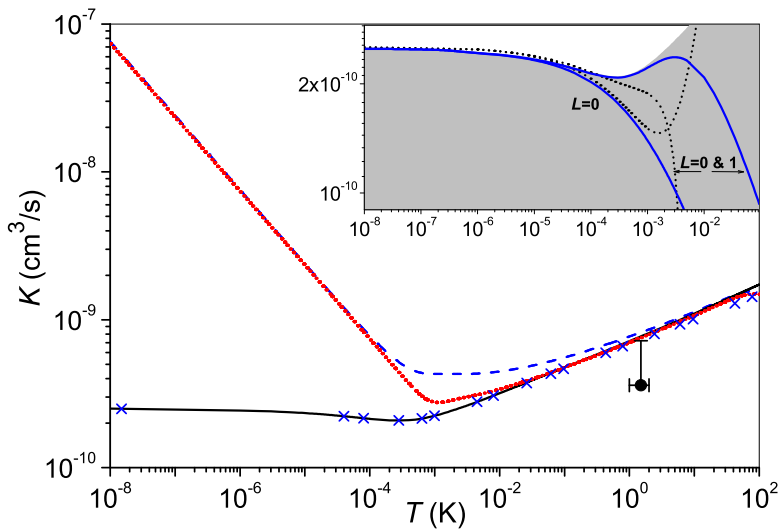}
\caption{Capture rate coefficients for the Li~+~CaH reaction. Quantum (solid line), classical (dotted line) and WKB-corrected (dashed line) results are compared with the experiment (dot) \cite{CaHLiJonathan2012} and quantum calculations with dispersion EP (crosses). In the inset, quantum capture rate coefficients computed with $L=0$ and $L=0,1$ contributions (solid lines) are compared with those given by the QL quantum defect model \cite{BoGao}. The shaded area represents the converged rate coefficient. } 
\centering
\label{RCfig}
\end{figure}

Figure~\ref{RCfig} compares our calculated capture rates with  the experimental result measured by Weinstein and co-workers \cite{CaHLiJonathan2012} at $T=1-2$ K ($K = 3.6\times 10^{-10}$ cm$^3$/s with a factor of two uncertainty). Our quantum AC calculations  give  $K = 7.2\times 10^{-10}$ cm$^3$/s, in excellent agreement with experiment.  The classical AC rate is almost the same, $K = 7.1\times 10^{-10}$~cm$^3$/s. Decomposing the rate coefficient into the partial wave contributions, we found that all partial waves with $L\le 18$ make a significant contribution to the reaction rate, with the maximum contribution of 10\% from  $L=6,7$.

As shown in Fig.~\ref{RCfig} and noted above, the unphysical behavior of the classical $s$-wave scattering cross section in the $E \to 0$ limit translates into the unphysical $T^{-1/2}$ increase of the rate coefficient as $T\to 0$. Quantum AC calculations provide a constant  limiting value of the rate coefficient in accordance with the Wigner threshold law.  The quantum result approaches the classical limit with increasing temperature, with the two limits hardly distinguishable from each other already at $10-50$ mK. The onsets of $L=1$ and $L=2$ classical scattering correspond to 14 and 50 mK, respectively, so the main source of the ``quantum effects'' is  due to the $L=0$ partial wave. On the other hand, the classical approximation is unable to describe the qualitative behavior of the reaction rate in the $s$-wave limit at $T<1$~mK. We note that the crossover from classical to quantum $s$-wave scattering regime, naturally set by matching the de Broiglie wavelength to the scattering length, occurs at much higher temperatures of the order 400 mK (as estimated for pure dispersion $R^{-6}$ potential with the  scattering length expression from Ref.~\cite{Julienne-MiesJOSA1989}). 

The contributions due to higher-order attractive long-range interaction terms manifest themselves at even higher temperatures: according to quantum calculations, the difference between the {\it ab initio} derived AC potential and dispersion EP approaches 10\% for the rate coefficient above 100 K.

In order to estimate the deviation from universal behavior, we follow Refs.~\cite{RackhamJCP03,Manolo2014} and assume  statistical decay of the reaction complex. The reaction probability can then be estimated using the phase space theory as $\kappa = N_p/(N_p + N_r)$, which is essentially the number of available product channels divided by the total number of available channels for both products and reactants at a given collision energy. The availability of a channel is determined by its energy, angular momentum and parity constrains. Strictly speaking, this expression should be used for scaling the capture probabilities at fixed $L$ and collision energy \cite{Manolo2014}, but we used it directly for the rate coefficient setting the collision energy to 1~cm$^{-1}$ (approx. 1.5 K) and computing the numbers of channels available at all total angular momenta $J=L$ up to 18. To calculate the internal energy of the LiH molecule, we used the Dunham expansion coefficients from Ref.~\cite{CHA86}. The statistical spin factor was taken as unity for reactant channels (one fourth of quadruply degenerated channels are available for the reaction on the singlet PES) and as two for product channels (due to the doubly degenerate state of the product Li atom). The value of $\kappa=0.9996$ obtained in this way indicates negligible deviation of the reaction rate ($\kappa K$) from the capture rate $K$. However, according to Ref.~\citenum{Manolo2014}, rigorous quantum scattering calculations may give larger deviations than predicted by simple statistical considerations. 
}

In the inset of Fig.~\ref{RCfig}, we compare our numerically calculated quantum AC capture rates with the analytical ``Quantum Langevin'' (QL) model by Bo Gao \cite{BoGao}. Our numerical results perfectly reproduce the zero-temperature limit and almost coincide with QL calculations  below $T = 0.1$ mK. The reason for the divergence at higher temperatures is the power series expansion in temperature used in the QL model, as can be inferred by comparing the rate coefficients computed with $L=0$ and $L = 0,1$ contributions. The excellent agreement between the QL and numerical results at $T \to 0$ confirms the high accuracy of the present calculations. 


\section{Summary and Conclusions}

We have applied the adiabatic capture theory to study the chemical reaction Li + CaH $\to$ LiH + Ca using an accurate {\it ab initio} PES of $^3A'$ symmetry  \cite{ourCaHLi}. The basic idea of the capture theories is to separate the configuration space of the reactants into the short-range and long-range regions. In the long-range region, the exchange interaction which determines the energy splitting between the triplet and the singlet PES, goes to zero exponentially with $R$ \cite{Smirnov,Stwalley}. The asymptotic behaviour of the singlet PES (on which the chemical reaction occurs)  is thus identical to that of the triplet PES. As a result, we can use the triplet PES developed previously \cite{ourCaHLi}  to compute the accurate ACs and capture probabilities for Li-CaH reactive scattering.

The AC potentials, which determine  the reaction probabilities in the capture approximation, were computed by diagonalizing the Li-CaH Hamiltonian in Jacobi coordinates. The ACs are parametrized by the orbital angular momentum of the reaction complex $L$ and have a centrifugal barrier for $L>0$ that prevents the reactants from reaching the short-range region where the reaction occurs. The capture probabilities are defined according to a simple classical Langevin prescription, Eq.~(\ref{Pclass}), that depends only on the barrier height, with a WKB correction applied to account for quantum tunnelling. This physical  scenario corresponds to the reactant flux being fully absorbed at short range and is represented by the boundary condition (\ref{WFtoR0}) applied at  $R=R_0$. The results for $\alpha(E)$ depend on $R_0$ only weakly, and we take an average over a range of $R_0$ to obtain the most probable estimate for the transmission coefficients. As shown in Figs. 2 and 3, the transmission coefficients for $L>0$ decline rapidly as collision energy decreases below the barrier height. The $L=0$ capture probability declines much more gradually  with collision energy due to the absence of the centrifugal barrier (Fig. 2).


The total reaction cross section summed over all partial waves is dominated by $s$-wave scattering at collision energies below 1 mK, where the use of the quantum capture theory \cite{Bohn10} is essential to reproduce the $s$-wave Wigner limit of the reaction rate.  At higher energies, more partial waves contribute and the total cross section and rate constant can be accurately described by the classical capture theory.  The calculated reaction rate agrees well with the experimental result at 1 K (Fig. 4)  \cite{CaHLiJonathan2012}, demonstrating that adiabatic capture theories can  predict the reaction rates with nearly quantitative  accuracy  in the multiple partial wave regime.

The adiabatic capture theories used here \cite{ourKRb,ClaryARPC90,TroeACP97,BoGao,Julienne10,Bohn10} offer a computationally efficient way of handling reactive  channels in barrierless chemical reactions that proceed through the formation of a deeply bound complex. These theories could be extended to  study  the   effects of intermolecular and intramolecular spin-dependent  interactions on chemical reactions involving molecular radicals such as CaH, OH, and NH.  The intramolecular spin-rotation interaction in CaH is weak, and hence expected to be of minor importance for Li~+~CaH; however, it may no longer be the case for reactions involving $^3\Sigma$ molecules like NH, in which the spin-spin interaction can be comparable to collision energy at 1 K. A model study of  the chemical reaction NH + NH found that spin-forbidden transitions between the PESs of different multiplicity can be efficient and reaction rates can be large even if the NH molecules are initially in the fully spin-polarized states \cite{NH}.  More sophisticated non-adiabatic capture theories developed by Rackham, Alexander, and Manolopoulos \cite{RackhamJCP03,RackhamJCP04} appear to be well-suited to describe these effects.

Another possibility is to combine the capture approximation for reaction probabilities with an accurate coupled-channel description of the Li-CaH complex in the entrance reaction channel. Such a combined theoretical approach would be extremely useful to study the interplay between inelastic and reactive  scattering, and to elucidate  the effects of external fields on chemical reactivity. It has been predicted that if open-shell atoms and molecules are spin-polarized prior to collision, they will typically not react, but rather scatter inelastically \cite{prl06,jcp07}. Flipping the electron spin of one (or both) of the reactants could  then be used to trigger chemical reactions in ultracold atom-molecule mixtures \cite{prl06,jcp07}. The feasibility of this control mechanism can be studied using the combined approach. 



\begin{acknowledgments}
We thank Prof. Jonathan Weinstein for fruitful discussions. This work was partially supported by the Russian Academy of Sciences (Program of the Fundamental Research by Division of Chemistry and Material Sciences 01 coordinated by Acad. O.M. Nefedov). 
\end{acknowledgments}


\begin{thebibliography}{99}

\bibitem{ColdMoleculeBook}
{\it Cold Molecules: Theory, Experiment, Applications} ed. R. V. Krems, W. C. Stwalley, and B. Friedrich, (CRC press, Taylor \& Francis, 2009).

\bibitem{molphys13}
M. Lemeshko, R. V. Krems, J. M. Doyle, and S. Kais, Mol. Phys. {\bf 111}, 1648 (2013).

\bibitem{njp09}
L. D. Carr, D. DeMille, R. V. Krems, and J. Ye, New J. Phys. {\bf 11}, 055049 (2009).

\bibitem{spectroscopy}
M. Quint\'ero-Perez, T. E. Wall, S. Hoekstra, and H. L. Bethlem, arXiv:1405.2751v1 (2014).

\bibitem{JunARPC}
B. K. Stuhl, M. T. Hummon, and J. Ye,  Annu. Rev. Phys. Chem. {\bf 65}, 501 (2014).

\bibitem{ChemRevBas}
S. Y. T. van de Meerakker, H. L. Bethlem, N. Vanhaecke,  and G. Meijer, Chem. Rev. {\bf 112}, 4828 (2012).

\bibitem{ChemRevJohn} 
N. R. Hutzler, H.-I. Lu, and J. M. Doyle,  Chem. Rev. {\bf 112}, 4803  (2012).

\bibitem{Roman10}
F. Herrera and R. V. Krems, \pra{84}{051401(R)}{2011}.

 \bibitem{Roman13}
 P. Xiang, M. Litinskaya, E. A. Shapiro, and R. V. Krems, New J. Phys. {\bf 15}, 063015 (2013).

\bibitem{John14}
Hsin-I Lu, I. Kozyryev, B. Hemmerling, J. Piskorski, and J. M. Doyle, \prl{112}{113006}{2014}.

\bibitem{KRb08}
K.-K. Ni, S. Ospelkaus, M. H. G. de Miranda, A. Pe'er, B. Neyenhuis, J. J. Zirbel, S. Kotochigova,  P. S. Julienne, D. S. Jin, and J. Ye,  Science {\bf 322}, 231 (2008).

\bibitem{KRb10a}
S. Ospelkaus, K.-K. Ni, D. Wang, M. H. G. de Miranda, B. Neyenhuis, G. Quemener, P. S. Julienne, J. L. Bohn, D. S. Jin, and J. Ye, Science {\bf 327}, 853 (2010).

\bibitem{KRb10b}
K.-K. Ni, S. Ospelkaus, D. Wang, G. Qu{\'e}m{\'e}ner, B. Neyenhuis, M. H. G. de Miranda, J. L. Bohn,  J. Ye, and D. S. Jin, Nature {\bf 464}, 1324 (2010).

\bibitem{ChemRevGoulven}
G. Qu{\'e}m{\'e}ner and P. S. Julienne,  Chem. Rev. {\bf 112}, 4949 (2012).

 \bibitem{prl06}
T. V. Tscherbul  and R. V. Krems, Phys. Rev. Lett. {\bf 97}, 083201 (2006).

\bibitem{jcp07}
E. Abrahamsson, T. V. Tscherbul, and R. V. Krems, J. Chem. Phys. {\bf 127}, 044302 (2007).
 
 \bibitem{Roman08}
  R. V. Krems,  Phys. Chem. Chem. Phys. {\bf 10}, 4079 (2008).
 
\bibitem{Pavel1}
M. T. Cvita\u{s}, P. Sold{\'a}n, J. M. Hutson, P. Honvault, and J.-M. Launay, Phys. Rev. Lett. {\bf 94}, 033201 (2005).

\bibitem{GoulvenK3}
G. Qu\'em\'ener, P. Honvault, J.-M. Launay, P. Sold\'an, D. E. Potter, and J. M. Hutson, Phys. Rev. A {\bf 71}, 032722 (2005).

\bibitem{ourLiHF}
T. V. Tscherbul and R. V. Krems,  arXiv:1409.5857 (2014).

\bibitem{Bohn10}
G. Qu{\'e}m{\'e}ner and J. L. Bohn,  Phys. Rev. A {\bf 81}, 022702 (2010).

\bibitem{Bohn12}
M. H. G. de Miranda, A. Chotia,  B. Neyenhuis, D. Wang, G. Qu{\'e}m{\'e}ner, S. Ospelkaus, J. L. Bohn, J. Ye, and D. S. Jin,  Nat. Phys. {\bf 7}, 502 (2011).

\bibitem{Julienne10}
Z. Idziaszek and P. S. Julienne, Phys. Rev. Lett. {\bf 104}, 113202 (2010).
\bibitem{Langevin} P. Langevin, Ann. Chem. Phys. {\bf 5}, 245 (1905). 
\bibitem{Julienne13}
K. Jachymski, M. Krych, P. S. Julienne, and Z. Idziaszek, \prl{110}{213202}{2013}.

\bibitem{Narevicius}
A. B. Henson, S. Gersten, Y. Shagam, J. Narevicius, and E. Narevicius, Science {\bf 338}, 234 (2012).


\bibitem{BoGao}
Bo Gao,  Phys. Rev. Lett. {\bf 105}, 263203 (2010).


\bibitem{ClaryARPC90}
D. C. Clary, Annu. Rev. Phys. Chem. {\bf 41}, 61 (1990).
\bibitem{TroeACP97}
J. Troe, Adv. Chem. Phys. {\bf 101}, 819 (1997). 
\bibitem{ClaryMP82} 
D. C. Clary, Mol. Phys. {\bf 48}, 619 (1982). 
\bibitem{RackhamJCP03} E. J. Rackham, T. Gonzales-Lezana, and D. E. Manolopoulos, J. Chem. Phys. {\bf 119}, 12895 (2003).
\bibitem{RackhamJCP04} M. H. Alexander, E. J. Rackham, and D. E. Manolopoulos, J. Chem. Phys. {\bf 121}, 5221 (2004)
\bibitem{VMU5} A. A. Buchachenko, Moscow Univ. Chem. Bull. {\bf 53}, 159 (2012).
\bibitem{ourKRb} A. A. Buchachenko, A. V. Stolyarov, M. M. Szcz\c{e}\'sniak, and G. Cha{\l}asi{\'n}ski,
J. Chem. Phys. {\bf 137}, 114305 (2012).


\bibitem{CaHLiJonathan2012} V. Singh, K. S. Hardman, M.-J. Lu, A. Ellis, M. J. Morrison, and J. D. Weinstein, Phys. Rev. Lett. {\bf 108}, 203201 (2012). 

\bibitem{ourCaHLi} T. V. Tscherbul, J. K{\l}os, and A. A. Buchachenko, Phys. Rev. A {\bf 84}, 040701 (2011). 

\bibitem{HuberHerzberg} K. P. Huber and G. Herzberg, {\it Molecular Spectra and Molecular Structure. IV. Constants of Diatomic Molecules} (Van Nostrand Reinhold: New York, 1979).
\bibitem{Smirnov}
B. M. Smirnov and M. I. Chibisov, Sov. Phys. JETP {\bf 21}, 624 (1965).

\bibitem{Stwalley}
W. T. Zemke and W. C. Stwalley, \jcp{111}{4962}{1999}.

\bibitem{BotalibGadea} A. Botalib and F. X. Gadea, J. Chem. Phys. {\bf 97}, 1144 (1992). 
\bibitem{Shayesteh} 
A. Shayesteh, K. A. Walker, I. Gordon, D. R. T. Appado, and P. F. Bernath, J. Mol. Struct. {\bf 695}, 23 (2004).
\bibitem{GDB} G. Delgado-Barrio and J. A. Beswick, in {\it Structure and Dynamics of Non-Rigid Molecular Systems}, Y.G. Smeyers (Ed.) (Kluwer: Dordrecht, 1994), p.203.
\bibitem{Nadine}
B. P. Reid, K. C. Janda, and N. Halberstadt, J. Phys. Chem. {\bf 92}, 587 (1988).
\bibitem{RamillonMcCarrol}
M. Ramillon and R. McCarroll, J. Chem. Phys. {\bf 101}, 8697 (1994).
\bibitem{TroeCPL85}
J. Troe, Chem. Phys. Lett. {\bf 122}, 425 (1985).
\bibitem{Vogt-Wannier} E. Vogt and G. E. Wannier, Phys. Rev. {\bf 95}, 1190 (1954). 
\bibitem{Light} J. C. Light and A. Altenberger-Siczek, J. Chem. Phys. {\bf 64}, 1907 (1976). 
 
 
\bibitem{Fernandez-Ramos} A. Fern{\'a}ndez-Ramos, J. A. Miller, S. J. Klippenstein, and D. G. Truhlar, Chem. Rev. {\bf 106}, 4518 (2006). 
\bibitem{OMalley} T. F. O'Malley, L. Spruch, and L. Rosenberg, J. Math. Phys. {\bf 2}, 491 (1961).
\bibitem{Watanabe} S. Watanabe and C. H. Greene, Phys. Rev. A {\bf 22}, 158 (1980). 
\bibitem{Cavangero} M. J. Cavagnero, Phys. Rev. A {\bf 50}, 2841 (1994). 
\bibitem{Gao1998} B. Gao, Phys. Rev. A. {\bf 58}, 1728 (1998). 
\bibitem{Gao2009} B. Gao, Phys. Rev. A. {\bf 80}, 012702 (2009). 
\bibitem{Michelli} A. Micheli, Z. Idziaszek, G. Pupillo, M. A. Baranov, P. Zoller, and P. S. Julienne, Phys. Rev. Lett. {\bf 105}, 073202 (2010). 
\bibitem{PP87} R. T. Pack, and G. A. Parker,   J. Chem. Phys. {\bf 87}, 3888 (1987).

\bibitem{LL} L. D. Landau and E. M. Lifshitz, {\it Quantum Mechanics} (Butterworth-Heinemann: Oxford, 2003).
\bibitem{Froman} N. Fr{\"o}man and P.-O. Fr{\"o}man, {\it JWKB Approximation: Contributions to the Theory} (North-Holland: Amsterdam, 1965).
\bibitem{TK} D. G. Truhlar and A. Kuppermann, J. Am. Chem. Soc. {\bf 93}, 1840 (1971). 
\bibitem{Klots1976} C. E. Klots, Chem. Phys. Lett. {\bf 38}, 61 (1976). 
\bibitem{Dashevskaya2003} E. I. Dashevskaya, A. I. Maergoiz, J. Troe, I. Litvin, and E. E. Nikitin,  J. Chem. Phys. {\bf 118}, 7313 (2003).
\bibitem{Nikitin2010} E. E. Nikitin and J. Troe, J. Phys. Chem. A {\bf 114}, 9762 (2010).
\bibitem{Julienne-MiesJOSA1989} P. S. Julienne and F. H. Mies, J. Opt. Soc. Am. {\bf 6}, 2257 (1989). 

\bibitem{Manolo2014} M. Lara, P. G. Jambrina, J.-M. Launay, and F. J. Aoiz, arXiv:1411.2666v1 (2014).
\bibitem{CHA86} Y. C. Chan, D. R. Harding, W. C. Stwalley, and C. R. Vidal, J. Chem. Phys. {\bf 85}, 2436 (1986). 

\bibitem{NH}
L. M. C. Janssen, A. van der Avoird,  and G. C. Groenenboom,  Phys. Rev. Lett. {\bf 110}, 063201 (2013).

\end{thebibliography}
\end{document}